# Scientific Case for Avoiding Dangerous Climate Change to Protect Young People and Nature

James Hansen[a,1,2,3], Pushker Kharecha[a], Makiko Sato[a], Frank Ackerman[b], Paul J. Hearty[c], Ove Hoegh-Guldberg[d], Shi-Ling Hsu[e], Fred Krueger[f], Camille Parmesan[g], Stefan Rahmstorf[h], Johan Rockstrom[i], Eelco J. Rohling[j], Jeffrey Sachs[k], Pete Smith[l], Konrad Steffen[m], Lise Van Susteren[n], Karina von Schuckmann[o], James C. Zachos[p],

[a] NASA Goddard Institute for Space Studies and Columbia University Earth Institute, New York, NY 10025, [b]Stockholm Environment Institute-US Center, Tufts University, Medford, MA, [c]Department of Environmental Studies, University of North Carolina at Wilmington, NC, [d]Global Change Institute, University of Queensland, St. Lucia, Queensland, Australia, [e]Faculty of Law, University of British Columbia, Canada, [f]National Religious Coalition on Creation Care, Santa Rosa, CA 95407-6828, [g]Integrative Biology, University of Texas, Austin, TX, and Marine Institute, University of Plymouth, UK, [h]Potsdam Institute for Climate Impact Research, Germany, [i]Stockholm Resilience Center, Stockholm University, Sweden, [j]School of Ocean and Earth Science, University of Southampton, United Kingdom, [k]Columbia University Earth Institute, New York, NY 10027, [l]University of Aberdeen, United Kingdom, [m]Cooperative Institute for Research in Environmental Sciences, University of Colorado, [n]Advisory Board, Center for Health and Global Environment, Harvard Medical School, [o]Centre National de la Recherche Scientifique, LOCEAN, Paris (hosted by Ifremer, Brest), France, [p]Earth and Planetary Science, University of California at Santa Cruz

**Summary.** Humanity is now the dominant force driving changes of Earth's atmospheric composition and thus future climate (1). The principal climate forcing is carbon dioxide ($CO_2$) from fossil fuel emissions, much of which will remain in the atmosphere for millennia (1, 2). The climate response to this forcing and society's response to climate change are complicated by the system's inertia, mainly due to the ocean and the ice sheets on Greenland and Antarctica. This inertia causes climate to appear to respond slowly to this human-made forcing, but further long-lasting responses may be locked in. We use Earth's measured energy imbalance and paleoclimate data, along with simple, accurate representations of the global carbon cycle and temperature, to define emission reductions needed to stabilize climate and avoid potentially disastrous impacts on young people, future generations, and nature. We find that global $CO_2$ emissions reduction of about 6%/year is needed, along with massive reforestation.

Governments have recognized the need to limit emissions to avoid dangerous human-made climate change, as formalized in the Framework Convention on Climate Change (3), but only a few nations have made substantial progress in reducing emissions. The stark reality (4) is that global emissions are accelerating and new efforts are underway to massively expand fossil fuel extraction, by oil drilling to increasing ocean depths, into the Arctic, and onto environmentally fragile public lands; squeezing of oil from tar sands and tar shale; hydro-fracking to expand extraction of natural gas; and increased mining of coal via mechanized longwall mining and mountain-top removal.

---

[1] Author contributions: J.H. conceived and drafted the paper, based on inputs from and multiple iterations with all co-authors; P.K. carried out carbon cycle calculations; P.K. and M.S. carried out temperature calculations; M.S. made all figures; Paul Epstein (deceased) drafted the health section.
[2] The authors declare no conflict of interest.
[3] To whom correspondence should be addressed: james.e.hansen@nasa.gov

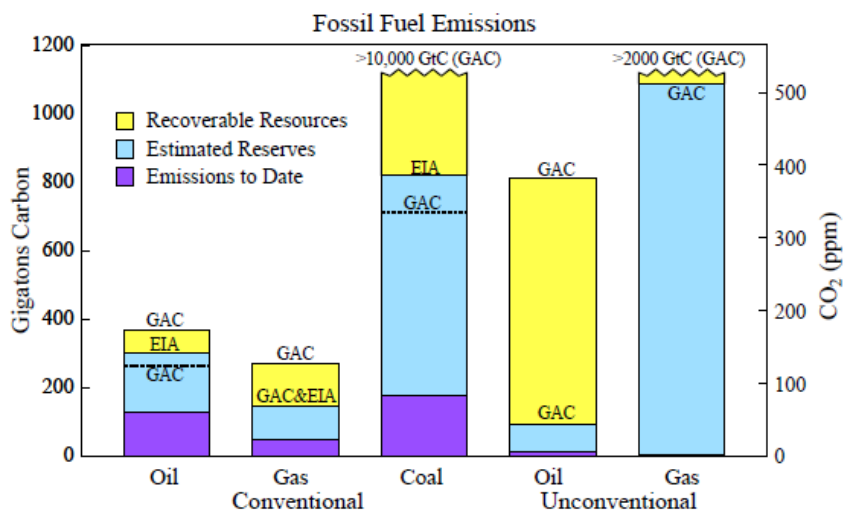

**Fig. P1.** $CO_2$ emissions by fossil fuels (1 ppm $CO_2$ ~ 2.12 GtC). Estimated reserves and potentially recoverable resources are from EIA (9) and GAC (10).

Governments not only allow this activity, but use public funds to subsidize fossil fuels at a rate of 400-500 billion US$ per year (5). Nor are fossil fuels required to pay their costs to society. Air and water pollution from extraction and burning of fossil fuels kills more than 1,000,000 people per year and affects the health of billions of people (6). But the greatest costs to society are likely to be the impacts of climate change, which are already apparent and are expected to grow considerably (7, 8).

Fossil fuel emissions to date are only a small fraction of potential emissions from known reserves and potentially recoverable resources (Fig. P1). Although there are uncertainties in reserves and resources, ongoing fossil fuel subsidies and continuing technological advances ensure that more and more of these fuels will be economically recoverable.

Burning all fossil fuels would create a very different planet than the one that humanity knows. The paleoclimate record and ongoing climate change make it clear that the climate system would be pushed beyond tipping points, setting in motion irreversible changes, including ice sheet disintegration with a continually adjusting shoreline, extermination of a substantial fraction of species on the planet, and increasingly devastating regional climate extremes.

Earth's paleoclimate history helps us assess levels of global temperature consistent with maintaining a planet resembling that to which civilization is adapted, for example, avoiding sea level rise of many meters. Earth's measured energy imbalance during a time of minimum solar irradiance, with Earth absorbing more solar energy than the heat energy it radiates to space, confirms the dominant effect of increasing atmospheric $CO_2$ on global temperature (11) and allows us to determine fossil fuel emission reductions needed to restore Earth's energy balance, which is the basic requirement for stabilizing climate.

We conclude that initiation of phase-out of fossil fuel emissions is urgent. For example, if emission reductions begin this year the required rate of decline is 6%/year to restore Earth's energy balance, and thus approximately stabilize climate, by the end of this century. If emissions reductions had begun in 2005, the required rate was 3%/year. If reductions are delayed until 2020, the required reductions are 15%/year. And these scenarios all assume a massive 100 GtC reforestation program, essentially restoring biospheric carbon content to its natural level.

The implication is that we must transition rapidly to a post-fossil fuel world of clean energies. This transition will not occur as long as fossil fuels remain the cheapest energy in a system that does not incorporate the full cost of fossil fuels. Fossil fuels are cheap only because they are subsidized, and because they do not pay their costs to society. The high costs to human health, food production, and natural ecosystems of air and water pollution caused by fossil fuel extraction and use are borne by the public. Similarly, costs of climate change and ocean acidification will be borne by the public, especially by young people and future generations.

Thus the essential underlying policy is for emissions of $CO_2$ to come with a price that allows these costs to be internalized within the economics of energy use. The price should rise over decades to enable people and businesses to efficiently adjust their lifestyles and investments to minimize costs.

Fundamental change is unlikely without public support. Gaining that support requires widespread recognition that a prompt orderly transition to the post fossil fuel world, via a rising price on carbon emissions, is technically feasible and may even be economically beneficial apart from the benefits to climate.

The most basic matter is not one of economics, however. It is a matter of morality – a matter of intergenerational justice. As with the earlier great moral issue of slavery, an injustice done by one race of humans to another, so the injustice of one generation to all those to come must stir the public's conscience to the point of action.

1. Intergovernmental Panel on Climate Change (IPCC), 2007: *Climate Change 2007: The Physical Science Basis*, Solomon, S.*, et al.* eds., Cambridge University Press, 996 pp.
2. Archer, D., 2005: Fate of fossil fuel $CO_2$ in geologic time. *J Geophy Res*, **110**, C09S05.
3. United Nations Framework Convention on Climate Change (FCCC), 1992: United Nations, http://www.unfccc.int.
4. Krauss, C., 2010: There will be fuel. *New York Times*, Page F1, New York edition, November 17, 2010.
5. G20 Summit Team, 2010: *Analysis of the Scope of Energy Subsidies and Suggestions for the G-20 Initiative*.
6. Cohen, A.J.*, et al.*, 2005: The Global Burden of Disease Due to Outdoor Air Pollution. *J Toxicol Environ Health, Part A*, **68**, 1301-1307.
7. Intergovernmental Panel on Climate Change (IPCC), 2007: *Climate Change 2007, Impacts, Adaptation and Vulnerability*, M.L. Parry, E. A. ed., Cambridge Univ Press, 996 pp.
8. Ackerman, F. and Stanton, E.A., 2011: Climate Risks and Carbon Prices: Revising the Social Cost of Carbon: http://www.economics-ejournal.org/economics/discussionpapers/2011-40 accessed Dec. 25, 2011.
9. Energy Information Administration (EIA), 2011: International Energy Outlook: http://www.eia.gov/forecasts/ieo/pdf/0484(2011).pdf accessed Sep 2011.
10. German Advisory Council on Global Change (GAC), 2011: World in Transition - A Social Contract for Sustainability.: http://www.wbgu.de/en/flagship-reports/fr-2011-a-social-contract/ accessed Oct 2011.
11. Hansen, J., Sato, M., Kharecha, P., and von Schuckmann, K., 2011: Earth's energy imbalance and implications. *Atmos Chem Phys*, **11**, 13421-13449.

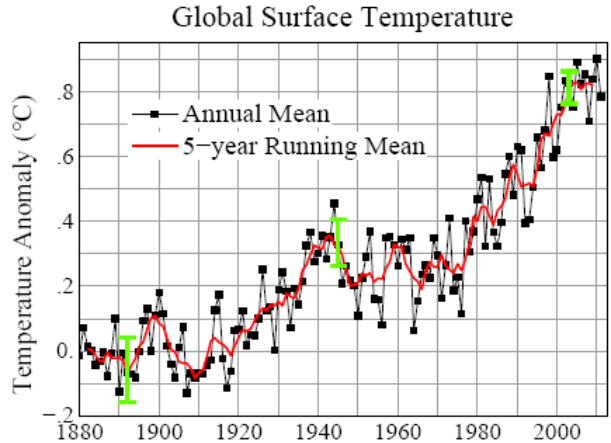

**Fig. 1.** Global surface temperature anomalies relative to 1880-1920 mean. Green bars are 95% confidence intervals (14).

**Global warming due to human-made gases, mainly $CO_2$, is already 0.8°C and deleterious climate impacts are growing worldwide. More warming is "in the pipeline" because Earth is out of energy balance, with absorbed solar energy exceeding planetary heat radiation. Maintaining a climate that resembles the Holocene, the world of relatively stable climate and shorelines in which civilization developed, requires rapidly reducing fossil fuel $CO_2$ emissions. Such a scenario is economically manageable and has multiple benefits for humanity and other species. Yet fossil fuel extraction is expanding, including highly carbon-intensive sources that can push the climate system beyond tipping points such that amplifying feedbacks drive further climate change beyond humanity's control. This situation raises profound moral issues in that young people, future generations, and nature, with no possibility of protecting their future well-being, will bear the principal consequences of actions and inactions of today's adults.**

We seek to clarify and quantify the urgency of phasing out fossil fuel emissions for the sake of avoiding disastrous climate change. We use Earth's paleoclimate history to determine the levels of global temperature that are consistent with maintaining a planet resembling that to which civilization is adapted. We use a tested carbon cycle model and a simple representation of global temperature and climate sensitivity consistent with paleoclimate data to determine the fossil fuel emission reductions that will be required to restore Earth's energy balance, which is the basic requirement for stabilizing climate. We also discuss the moral issues, our obligations to young people, future generations, less developed nations, indigenous people, and our fellow species.

## Global Temperature

Global surface temperature fluctuates stochastically and also responds to natural and human-made climate forcings. Forcings are imposed perturbations of Earth's energy balance such as changes of the sun's luminosity and human-made increase of atmospheric $CO_2$.

**Modern Temperature.** Temperature change in the past century (Fig. 1) includes unforced variability and forced climate change. Unusual global warmth in 1998 was due to the strongest El Niño of the century, a temporary warming in the tropics caused by an irregular oscillation of the tropical ocean-atmosphere system. Cooling in 1992-1993 was due to stratospheric aerosols from the Mount Pinatubo volcanic eruption, which reduced sunlight reaching Earth's surface as much as 2%. The long-term global warming trend is predominately a forced climate change caused by increased human-made atmospheric gases, mainly $CO_2$ (1).

The basic physics underlying this global warming, the greenhouse effect, is simple. An increase of gases such as $CO_2$ has little effect on incoming sunlight but makes the atmosphere more opaque at infrared wavelengths that radiate heat to space. The resulting Earth energy imbalance, absorbed solar energy exceeding heat emitted to space, causes the planet to warm.

Efforts to assess dangerous climate change have focused on estimating a permissible level of global warming. The Intergovernmental Panel on Climate Change (1, 2) summarized broad-based assessments with a "burning embers" diagram, which indicated that major problems begin with global warming of 2-3°C. A probabilistic analysis (3), still partly subjective, found a median "dangerous" threshold of 2.8°C, with 95% confidence that the dangerous threshold was 1.5°C or higher. These assessments were relative to global temperature in 2000; add 0.7°C to obtain warming relative to 1880-1920. The conclusion that humanity could tolerate global warming up to a few degrees Celsius meshed with common sense. After all, people readily tolerate much larger regional and seasonal climate variations.

The fallacy of this logic emerged in recent years as numerous impacts of global warming became apparent. Summer sea ice cover in the Arctic plummeted in 2007 and 2011 to an area 40 percent less than a few decades earlier and Arctic sea ice thickness declined a factor of four faster than simulated in IPCC climate models (4). The Greenland and Antarctic ice sheets began to shed ice at a rate, now several hundred cubic kilometers per year, which is continuing to accelerate (5, 6). Mountain glaciers are receding rapidly all around the world with effects on seasonal freshwater availability of major rivers (7, 8). The hot dry subtropical climate belts have expanded as the troposphere has warmed and the stratosphere cooled (9-11), probably contributing to observed increases in the area and intensity of wildfires (12). The abundance of reef-building corals is decreasing at a rate of 0.5-2%/year, at least in part due to ocean warming and acidification caused by rising dissolved $CO_2$ (13-15). More than half of all wild species have shown significant changes in where they live and in the timing of major life events (16, 17). Mega-heatwaves, such as those in the Moscow area in 2010 and Texas in 2011, have become more widespread with the increase demonstrably linked to global warming (18).

In recognition of observed growing climate impacts while global warming is less than 1°C, reassessment of the dangerous level of warming is needed. Earth's paleoclimate history provides a valuable tool for that purpose.

**Paleoclimate Temperature.** Global surface temperature in the Pliocene and Pleistocene (Fig. 2) is inferred from the composition of shells of deep-sea-dwelling microscopic animals preserved in ocean sediments (19, 20). Surface temperature change between the Holocene and the last ice age was about 1.5 times larger than deep ocean temperature change, because deep ocean temperature was limited as it approached the freezing point (21). We assume the same surface temperature amplification toward higher temperatures, which thus may tend to overestimate Pliocene surface temperature (21).

We concatenate paleoclimate (Fig. 2) and modern (Fig. 1) temperature records via the assumption that peak Holocene temperature (prior to the warming of the past century) was +0.5° ±0.25°C relative to the 1880-1920 mean. The facts that Antarctica and Greenland are losing mass at an accelerating rate (5, 6) and sea level is rising at a rate (+3m/millennium) much higher than during the past several thousand years provide strong evidence that the temperature in the past decade (+0.75°C relative to 1880-1920) exceeded the prior Holocene maximum.

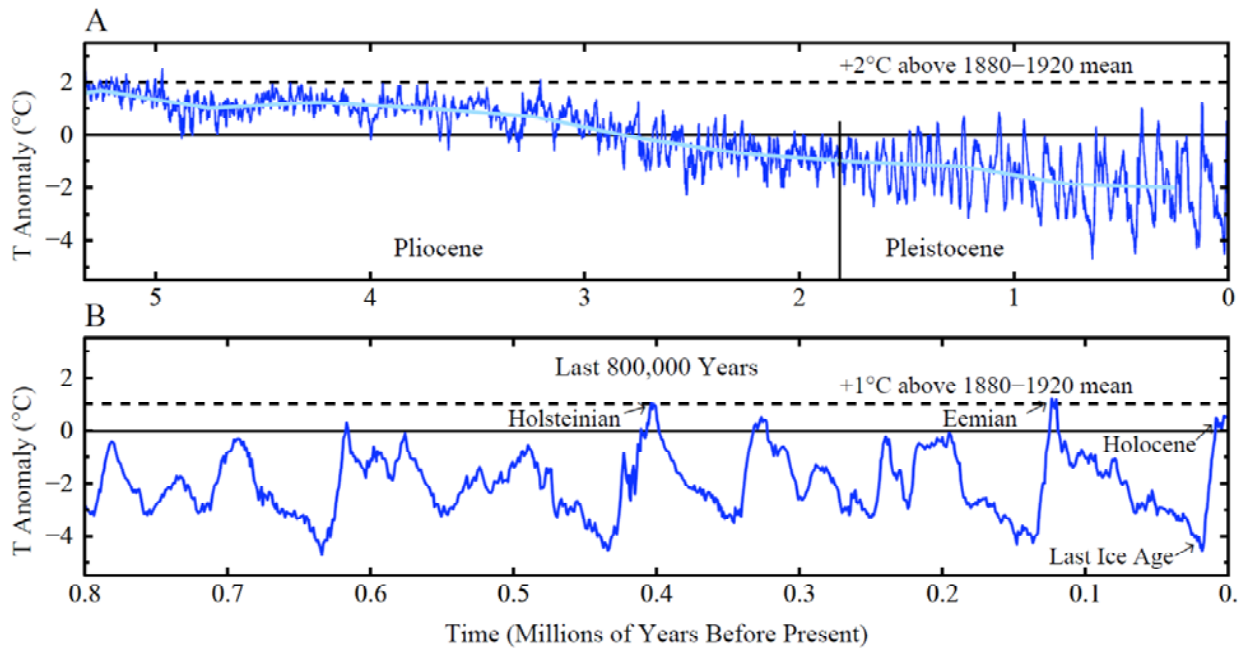

**Fig. 2.** Global temperature relative to 1880-1920 in (A) past 5,300,000 and (B) past 800,000 years (32).

Climate oscillations evident in Fig. 2 were instigated by small perturbations of Earth's orbit and the tilt of its spin axis relative to the orbital plane that alter the seasonal and geographical distribution of sunlight on the planet (19). These forcings change very slowly, with periods between 20,000 and 400,000 years, and thus the climate is able to stay in quasi-equilibrium with the forcings. The slow insolation changes instigated the climate oscillations in Fig. 2, but the mechanisms that caused the climate changes to be so large were two powerful amplifying feedbacks: the planet's surface albedo (its reflectivity, literally its whiteness) and the atmospheric $CO_2$ amount. As the planet warms, ice and snow melt, causing the surface to be darker, absorb more sunlight and warm further. As the ocean and soil become warmer they release $CO_2$ and other greenhouse gases, causing further warming. These amplifying feedbacks were responsible for almost the entire glacial-to-interglacial temperature change (22-24).

Albedo and $CO_2$ feedbacks acted as slaves to weak orbital forcings in the natural climate variations in Fig. 2, changing slowly over millennia. Today, however, $CO_2$ is under the control of humanity as fossil fuel emissions overwhelm natural changes. Atmospheric $CO_2$ has increased rapidly to a level not seen for at least 3 million years (25). Global warming induced by increasing $CO_2$ will cause ice to melt and sea level to rise as the global volume of ice moves toward the quasi-equilibrium amount that exists for a given global temperature. As the ice melts and its area decreases the albedo feedback will amplify global warming.

Paleoclimate data yield an estimate of eventual ice melt and sea level rise for a given global warming. The Eemian and Hosteinian interglacial periods (Fig. 2B), also known as marine isotope stages 5e and 11, respectively about 130,000 and 400,000 years ago, were about 1°C warmer than the 1880-1920 mean (Fig. 2B), where paleo temperature relative to the modern era is based on our conclusion above that peak global Holocene temperature exceeded the 1880-1920 mean by $0.5 \pm 0.25$°C. These prior two interglacials were warm enough for sea level to reach levels at least 4-6 meters higher than today (26-28). Ominously, global mean temperature 2°C higher than the 1880-1920 mean has not existed since at least the early to mid Pliocene (Fig. 2A). Inference of Pliocene sea level change from shoreline features is complicated by local tectonics, local sediment loading, convective flow in Earth's mantle, and regional vertical movement of the crust due to ice sheet loading or unloading (29), but the data suggest that sea level reached heights as much as 15-25 meters greater than today (29-32).

Paleoclimate records are less useful for estimating how fast ice sheets will respond to global warming, because the human-made climate forcing is nearly instantaneous compared with the slowly changing forcings that drove the climate changes in Fig. 2. However, paleoclimate data commonly exhibit sea level change of more than 1 m/century in response to climate forcing much smaller than the forcing that will occur this century with continuing fossil fuel use (26-28).

Global observations of on-going climate system changes provide another assessment tool. The rapid warming of the past three decades (Fig. 1) is already producing measurable effects, as discussed below.

## Earth's Energy Imbalance

At a time of climate stability, Earth radiates as much energy to space as it absorbs from sunlight. Today Earth is out of balance because increasing atmospheric gases such as $CO_2$ reduce Earth's heat radiation to space, causing an energy imbalance, more energy coming in than going out. This imbalance causes Earth to warm and move back toward energy balance, but warming and restoration of energy balance are slowed by Earth's thermal inertia, due mainly to the ocean.

The immediate planetary energy imbalance caused by a $CO_2$ increase can be calculated precisely. The radiation physics is rigorously understood and does not require a climate model. But the ongoing energy imbalance is reduced by the fact that Earth has already warmed 0.8°C, thus increasing heat radiation to space. The imbalance is also affected by other factors that alter climate, such as changes of solar irradiance, the reflectivity of Earth's surface, and aerosols.

Determination of the state of Earth's climate therefore requires measuring the energy imbalance. This is a challenge, because the imbalance is expected to be only about 1 $W/m^2$ or less, so accuracy approaching 0.1 $W/m^2$ is needed. The most promising approach is to measure the rate of changing heat content of the ocean, atmosphere, land, and ice (33).

**Observed Energy Imbalance.** Nations of the world have launched a cooperative program to measure changing ocean heat content, distributing more than 3000 Argo floats around the world ocean, with each float repeatedly lowering an instrument package to a depth of 2 km and back (34). Ocean coverage by floats reached 90% by 2005 (34), with the gaps mainly in sea ice regions, yielding the potential for an accurate energy balance assessment, provided that several systematic measurement biases exposed in the past decade are minimized (35, 36).

Analysis of the Argo data yields a heat gain in the ocean's upper 2000 m of 0.41 $W/m^2$ averaged over Earth's surface during 2005-2010 (37). Smaller contributions to planetary energy imbalance are from heat gain by the deeper ocean (+0.10 $W/m^2$), energy used in net melting of ice (+0.05 $W/m^2$), and energy taken up by warming continents (+0.02 $W/m^2$). Data sources for these estimates and uncertainties are provided elsewhere (33). The resulting net planetary energy imbalance for the six years 2005-2010 is +0.58 ±0.15 $W/m^2$.

This positive energy imbalance in 2005-2010 demonstrates that the effect of solar variability on climate is much less than the effect of human-made greenhouse gases. If the sun were the dominant forcing, the planet would have a negative energy balance in 2005-2010, when solar irradiance was at its lowest level in the period of accurate data, i.e., since the 1970s (38). Even though much of the greenhouse gas forcing has been expended in causing observed 0.8°C global warming, the residual positive forcing overwhelms the negative solar forcing, yielding a net planetary energy imbalance +0.58 ±0.15 $W/m^2$.

Earth's energy imbalance averaged over the 11-year cycle of solar variability should be larger than the measured +0.58 W/m$^2$ at solar minimum. The mean imbalance averaged over the solar cycle is estimated to be +0.75 ±0.25 W/m$^2$ (33).

**Implications for $CO_2$ Target.** Earth's energy imbalance is the single most vital number characterizing the state of Earth's climate. It informs us about the global temperature change "in the pipeline" without further change of climate forcings. It also defines how much we must reduce greenhouse gases to restore energy balance and stabilize climate, if other forcings remain unchanged. The measured energy imbalance accounts for all natural and human-made climate forcings, including changes of Earth's surface and atmospheric aerosols.

If Earth's mean energy imbalance is +0.5 W/m$^2$, $CO_2$ must be reduced from the current level of 390 ppm to about 360 ppm to increase Earth's heat radiation to space by 0.5 W/m$^2$ and restore energy balance. If Earth's energy imbalance is 0.75 W/m$^2$, $CO_2$ must be reduced to about 345 ppm to restore energy balance (33, 39).

The measured energy imbalance affirms that a good initial $CO_2$ target to stabilize climate near current temperatures is "<350 ppm" (20). Specification of a more precise $CO_2$ target now is difficult and unnecessary, because of uncertain future changes of other forcings including other gases, ground albedo, and aerosols. More precise knowledge of the best target will become available during the time that it takes to turn around $CO_2$ growth and approach the initial 350 ppm target.

Ironically, future reductions of particulate air pollution may exacerbate global warming by reducing the cooling effect of reflective aerosols. However, a concerted effort to reduce non-$CO_2$ forcings by methane, tropospheric ozone, other trace gases and black soot might counteract the warming from a decline in reflective aerosols (39). Our calculations below of future global temperature assume that compensation. If that goal is not achieved, future warming could exceed calculated values.

## Carbon Cycle and Atmospheric $CO_2$

The carbon cycle defines the fate of $CO_2$ injected into the air by fossil fuel burning (1, 40) as the $CO_2$ distributes itself over time among surface carbon reservoirs: the atmosphere, ocean, soil, and biosphere. We use the dynamic-sink pulse-response function version of the well-tested Bern carbon cycle model (41), as described elsewhere (20, 42).

A pulse of $CO_2$ injected into the air decays by half in about 25 years (Fig. 3), but nearly one-fifth is still in the atmosphere after 500 years. Eventually, over millennia, weathering of rocks will deposit this excess $CO_2$ on the ocean floor as carbonate sediments.

A negative $CO_2$ pulse decays at about the same rate as a positive pulse (Fig. 3A), which is an important fact for policy considerations. If it is decided in the future that $CO_2$ must be sucked from the air and removed from the carbon cycle (e.g., by making carbonate bricks or storing the $CO_2$ in underground reservoirs), the effect of the atmospheric $CO_2$ reduction will decline as the negative $CO_2$ increment becomes spread among the carbon reservoirs. The main process limiting the long-term impact on atmospheric $CO_2$ is degassing from the ocean as equilibrium is established among the carbon reservoirs.

How fast atmospheric $CO_2$ declines if fossil fuel emissions are instantly terminated (Fig. 3B) is instructive. Halting emissions in 2015 causes $CO_2$ to decline to 350 ppm at the century's end (Fig. 3b). A 20 year delay in halting emissions has $CO_2$ returning to 350 ppm at about 2300. With a 40 year delay, $CO_2$ does not return to 350 ppm until after 3000. These results show how difficult it is to get back to 350 ppm if high emissions continue for even a few decades.

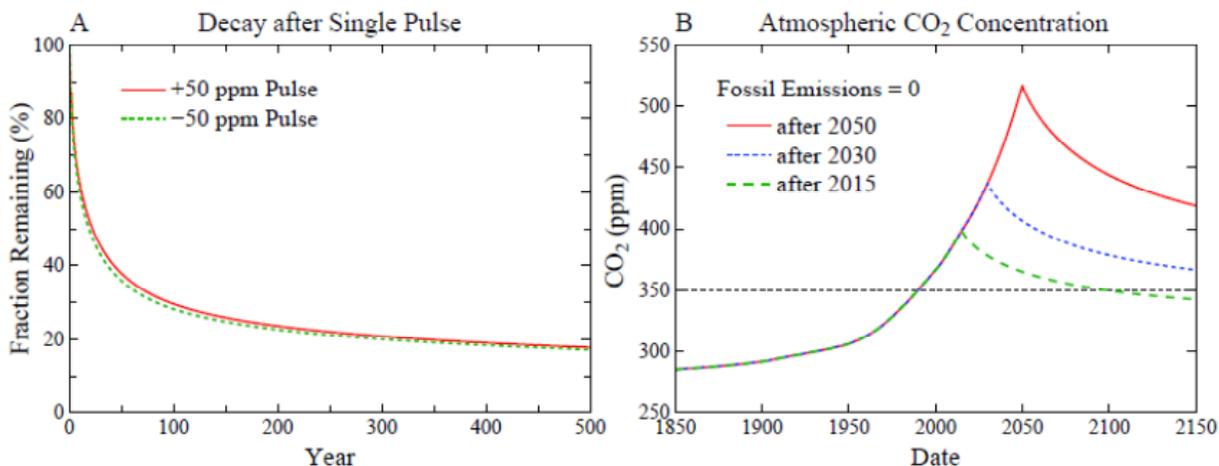

**Fig. 3.** (A) Decay of instantaneous injection or extraction of atmospheric $CO_2$, (B) $CO_2$ amount if fossil fuel emissions are terminated at the end of 2015, 2030, or 2050. Land use emissions terminate at the end of 2015 in all three cases.

**Reforestation and Soil Carbon.** The long $CO_2$ lifetime does not make it impossible to return $CO_2$ to 350 ppm this century. Reforestation and increasing soil carbon can help draw down atmospheric $CO_2$, even though the effect on atmospheric $CO_2$ amount decays (Fig. 3a).

Fossil fuels account for about 80% of the $CO_2$ increase from preindustrial 275 ppm to 390 ppm today, with deforestation accounting for the other 20%. Net deforestation to date is estimated to be 100 GtC (gigatons of carbon) with ±50% uncertainty (43).

Although complete restoration of deforested areas is unrealistic, a 100 GtC carbon storage is conceivable because: (1) the human-enhanced atmospheric $CO_2$ level increases carbon uptake by vegetation and soils, (2) improved agricultural practices can convert agriculture from a $CO_2$ source into a $CO_2$ sink (44), (3) biomass-burning power plants with $CO_2$ capture and storage can contribute to $CO_2$ drawdown.

Forest and soil storage of 100 GtC is a challenge, but it has other benefits. Reforestation has been successful in diverse places (45). Minimum tillage with biological nutrient recycling, as opposed to plowing and chemical fertilizers, could sequester 0.4-1.2 GtC/year (46, 47) while conserving water in soils, building agricultural resilience to climate change, and increasing productivity especially in smallholder rain-fed agriculture, thereby reducing expansion of agriculture into forested ecosystems (36, 48).

Reforestation may be most beneficial in the tropics (49, 50), avoiding potential unintended impacts of major reforestation elsewhere (51). Net deforestation in recent decades has occurred mostly in the tropics (1, 52), so a large amount of land suitable for reforestation (meeting UNFCCC criteria) exists there (53).

Use of bioenergy to draw down $CO_2$ should employ feedstocks from residues, wastes, and dedicated energy crops that do not compete with food crops, thus avoiding loss of natural ecosystems and cropland (54-56). Reforestation competes with agricultural land use; land needs could decline by reducing use of animal products, as livestock now consume more than half of all crops (57, 58).

Our reforestation scenarios decrease today's net deforestation rate linearly to zero in 2030, followed by a sinusoidal 100 GtC biospheric carbon storage over 2031-2080. Alternative timings do not alter conclusions about the potential to achieve a given $CO_2$ level such as 350 ppm.

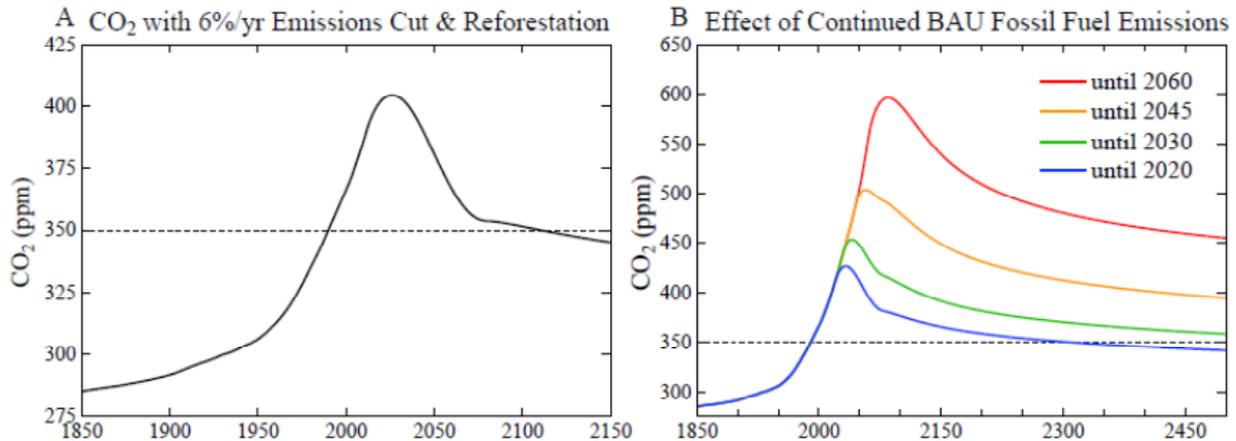

**Fig. 4.** (A) Atmospheric $CO_2$ if fossil fuel emissions are cut 6%/year beginning in 2013 and 100 GtC reforestation drawdown occurs in 2031-2080, (B) effect of delaying onset of emissions reduction.

**$CO_2$ Emission Reduction Scenarios.** A 6%/year decrease of fossil fuel emissions beginning in 2013, with 100 GtC reforestation, achieves a $CO_2$ decline to 350 ppm near the end of this century (Fig. 4A). Cumulative fossil fuel emissions in this scenario are ~136 GtC from 2012 to 2050, with an additional 15 GtC by 2100. If our assumed land use changes occur a decade earlier, $CO_2$ returns to 350 ppm several years earlier, however that has negligible effect on the global temperature maximum calculated below.

Conversely, delaying fossil fuel emission cuts until 2020 (with 2%/year emissions growth in 2012-2020) causes $CO_2$ to remain in the dangerous zone (above 350 ppm) until 2300 (Fig. 4B). If reductions are delayed until 2030, $CO_2$ remains above 400 ppm until almost 2500.

*These results emphasize the urgency of initiating emissions reduction.* If emissions reduction had begun in 2005, reduction at 3.5%/year would have achieved 350 ppm at 2100. Now the requirement is at least 6%/year. If we assume only 50 GtC reforestation, the requirement becomes at least 9%/year. Further delay of emissions reductions until 2020 requires a reduction rate of 15%/year to achieve 350 ppm in 2100.

**Geo-Engineering Atmospheric $CO_2$.** Perceived political difficulties of phasing out fossil fuel emissions have caused a surge of interest in possible "geo-engineering" designed to minimize human-made climate change (59). Such efforts must remove atmospheric $CO_2$, if they are to address direct $CO_2$ effects such as ocean acidification as well as climate change.

At present there are no technologies capable of large-scale air capture of $CO_2$. Keith et al. (60) suggest that, with strong research and development support and industrial scale pilot projects sustained over decades, costs as low as ~$500/tC may be achievable. An assessment by the American Physical Society (61) argues that the lowest currently achievable cost, using existing approaches, is much greater ($600/tCO_2$ or $2200/tC

The cost of removing 50 ppm of $CO_2$, at $500/tC, is ~$50 trillion (1 ppm $CO_2$ is ~2.12 GtC), but more than $200 trillion for the price estimate of the American Physical Society study. Moreover, the resulting atmospheric $CO_2$ reduction is only ~15 ppm after 100 years, because the extraction induces counteracting changes in the other surface carbon reservoirs – mainly $CO_2$ outgassing from the ocean (Fig. 3A). The estimated cost of maintaining a 50 ppm reduction on the century time scale is thus ~$150-600 trillion. The cost of air capture and storage of $CO_2$ may decline, but the practicality of carrying out such a program in response to a climate emergency is dubious, and today's young people and future generations would inherit a huge burden.

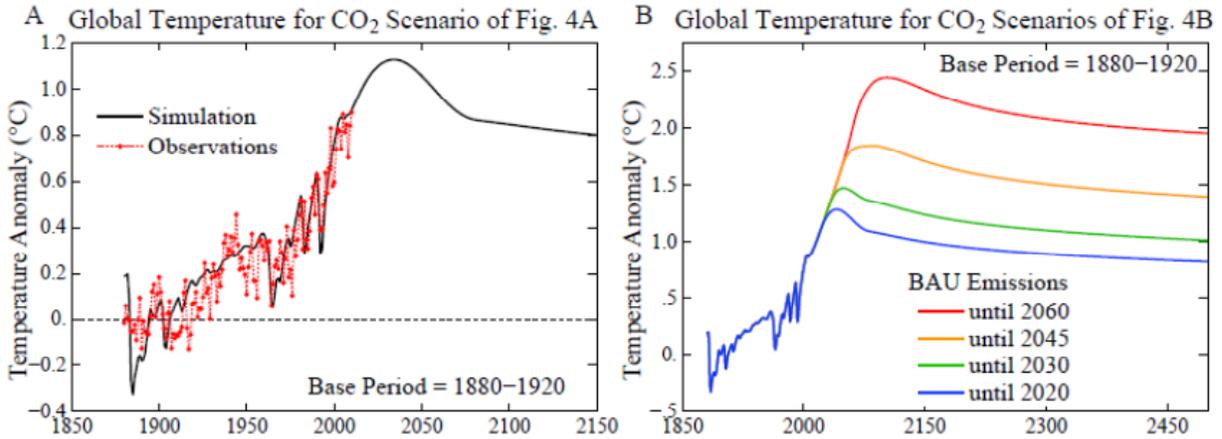

**Fig. 5.** Simulated global temperature relative to 1880-1920 mean for $CO_2$ scenarios of Fig. 4.

## Future Global Temperature Change

Future global temperature change depends primarily on atmospheric $CO_2$. $CO_2$ accounts for more than 80% of the growth of greenhouse gas climate forcing in the past 15 years (62). We approximate the net future change of human-made non-$CO_2$ forcings as zero, as discussed above. We neglect future changes of natural climate forcings, such as solar irradiance and volcanic aerosols, whose variability contributes little to long-term global temperature trend.

**Simulated Global Temperature.** We calculate global temperature change for a given $CO_2$ scenario using a climate response function that accurately replicates results from a global climate model with sensitivity 3°C for doubled $CO_2$ (33). Climate forcings that we use for the past [Fig. 4 of (62)] are updated annually at http://www.columbia.edu/~mhs119/GHG_Forcing/.

Simulated global temperature is shown in Fig. 5 for the $CO_2$ scenarios of Fig. 4. Peak global warming is ~1.1°C, declining to less than 1°C by mid-century, if $CO_2$ emissions are reduced 6%/year beginning in 2013. In contrast, warming reaches 1.5°C and stays above 1°C until after 2400 if emissions continue to increase until 2030, even though fossil fuel emissions are phased out rapidly (5%/year) after 2030 and 100 GtC reforestation occurs in 2031-2080. If fossil fuel emissions continue to increase until 2050, simulated global warming exceeds 2°C.

**Slow Climate Feedbacks and Tipping Points.** Our climate simulations, as with most climate models, incorporate only the effect of fast feedbacks in the climate system, such as water vapor, clouds, aerosols, and sea ice. Slow feedbacks, such as ice sheet disintegration are not included.

Excluding slow feedbacks is appropriate for the past century, because we know the ice sheets were stable and our climate simulations employ observed greenhouse gas amounts that include any changes caused by slow feedbacks. Exclusion of slow feedbacks in the 21$^{st}$ century, however, is a dubious assumption, which we used only because the rate at which slow feedbacks will come into play is unknown.

Slow feedbacks are important because of their impact on threshold or "tipping point" events (20, 63). Climate tipping points occur when climate change reaches a level where further large and possibly rapid changes become inevitable, proceeding mostly under their own momentum. Ice sheets provide an example. Once disintegration of an ice sheet is well underway the dynamics of the process takes over. At that point, reducing greenhouse gases cannot prevent substantial sea level rise.

Extermination of species can also reach a tipping point, because of interdependencies among species. Thus climate change that is large enough or fast enough can cause entire ecosystems to collapse, leading to mass extinctions (17).

Methane hydrates – methane molecules trapped in frozen water molecule cages in tundra and on continental shelves (64) – provide another potential tipping point. If methane hydrates thaw on a large scale they could greatly amplify global warming (65), because methane is a powerful greenhouse gas. There is already evidence of methane release from thawing permafrost on land (66) and from sea-bed deposits including methane hydrates (67).

**Dangerous Global Warming.** Tipping points help define the dangerous level of global warming, even though their nonlinear nature inhibits accurate predictability of temporal details of collapse. Assessment of tipping point threats is aided by the combination of paleoclimate records defining a quasi-equilibrium response to climate change and observations of ongoing dynamical responses to current global warming.

Ice sheets and sea level are a prime example. Paleoclimate data indicate that 1°C global warming above preindustrial levels (to the Eemian level) is likely to cause eventual sea level rise of several meters and 2°C (early Pliocene level) could cause eventual sea level rise as great as 15-25 m (29-32). Satellite measurements of Earth's gravity field reveal that Greenland and Antarctica are losing mass and the rate of loss has accelerated since measurements began in 2002 (5, 6), even though global temperature has barely risen above the Holocene temperature range in which the ice sheets have been stable for millennia.

Methane hydrates have been implicated by paleoclimate data as a likely principal mechanism in several rapid global warmings (68, 69). This appears to have occurred as a powerful feedback amplifying a natural warming trend (68-70), (see Supporting Information). Global warming of 2°C, amplified at high latitudes, would commit large areas of permafrost to thawing and might destabilize methane hydrates in ocean sediments.

Global warming to date is at most a few tenths of a degree above the prior Holocene range. Impacts on ice sheets and permafrost carbon are small so far, suggesting that these feedbacks may not be a major factor if global warming, now about 0.8°C, reaches a maximum of only ~1°C and then recedes, as in the scenario of Figs. 4A and 5A.

In contrast, the scenarios that reach 2°C or even 1.5°C global warming via only fast feedbacks appear to be exceedingly dangerous. These scenarios run a high risk of the slow feedbacks coming into play in major ways. However, we lack knowledge of how fast the slow feedbacks would occur, and thus which generations would suffer the greatest consequences.

The available information suggests that humanity faces a dichotomy of possible futures. Either we achieve a scenario with declining emissions, preserving a planetary climate resembling the Holocene, or the climate is likely to pass tipping points with amplifying feedbacks that assure transition to a very different planet with both foreseeable and unforeseen consequences.

## Likely Impacts of Global Warming

**Sea Level.** The prior interglacial period, the Eemian, was at most ~1°C warmer than the Holocene (Fig. 2). Sea level reached heights several meters above today's level with instances of sea level change by 1-2 m/century (26, 71). Geologic shoreline evidence has been interpreted as indicating a rapid sea level rise to a peak 6-9 meters above present late in the Eemian followed by a precipitous sea level fall (28, 72), although there remains debate within the research community about this specific history. An important point is that Eemian sea level excursions imply that rapid partial melting of Antarctic and/or Greenland ice occurred when the world was little warmer than today.

During the early Pliocene, which was only 1-2°C warmer than the Holocene (Fig. 2), sea level attained heights as much as 15-25 meters higher than today (29-32). Such sea level rise suggests that parts of East Antarctica must be vulnerable to eventual melting with global temperature increase of 1-2°C. Indeed, satellite gravity data and radar altimetry reveal that the Totten Glacier of East Antarctica, which fronts a large ice mass grounded below sea level, is already beginning to lose mass (73).

Expected human-caused sea level rise is controversial because predictions focus on sea level rise at a specific time, 2100. Prediction of sea level on a given date is inherently difficult, because it depends on how rapidly non-linear ice sheet disintegration begins. Focus on a single date also encourages people to take the estimated result as an indication of what humanity faces, thus failing to emphasize that the likely rate of sea level rise immediately after 2100 will be far larger than within the $21^{st}$ century, if $CO_2$ emissions continue to increase.

Most recent estimates of sea level rise by 2100 have been of the order of 1m, notably higher than estimates in earlier assessments (74), and it also has been argued (74, 75) that continued business-as-usual $CO_2$ emissions could cause multi-meter sea level rise this century. In Supplementary Material we discuss and provide references for estimated sea level rise and for observational evidence about changing ice sheet conditions.

The important point is that the uncertainty is not about whether continued rapid $CO_2$ emissions would cause large sea level rise – it is about how soon the large changes would begin. If all or most fossil fuels are burned, the carbon will remain in the climate system for many centuries, in which case multi-meter sea level rise is practically certain. Such a sea level rise would create hundreds of millions of global warming refugees from highly-populated low-lying areas, thus likely causing major international conflicts.

**Shifting Climate Zones.** Theory and climate models indicate that the tropical overturning (Hadley) atmospheric circulation expands poleward with global warming (9). There is evidence in satellite and radiosonde data and in reanalyses output for poleward expansion of the tropical circulation by as much as a few degrees of latitude since the 1970s (10, 11), which likely contributes to expansion of subtropical conditions and increased aridity in the southern United States (7, 76), the Mediterranean region, and southern Australia. Increased aridity and temperature contribute to increased forest fires that burn hotter and are more destructive (12).

Despite large year-to-year variability of temperature, decadal averages reveal isotherms (lines of a given average temperature) moving poleward at a typical rate of the order of 100 km/decade in the past three decades (77), although the range shifts for specific species follow more complex patterns (78). This rapid shifting of climatic zones far exceeds natural rates of change. Movement has been in the same direction (poleward, and upward in elevation) since about 1975. Wild species have responded to climate change, with at least 52 percent of species having shifted their ranges poleward as much as 600 km [and upward as much as 400 m (79)] in terrestrial systems and 1000 km in marine systems (16, 80).

Humans may adapt to shifting climate zones better than many species. However, political borders can interfere with human migration, and indigenous ways of life already have been adversely affected (74). Impacts are apparent in the Arctic, with melting tundra, reduced sea ice, and increased shoreline erosion. Effects of shifting climate zones also may be important for indigenous Americans who possess specific designated land areas, as well as other cultures with long-standing traditions in South America, Africa, Asia and Australia.

**Extermination of Species.** Biodiversity is affected by many agents including overharvesting, introduction of exotic species, land use changes, nitrogen fertilization, and direct effects of increased atmospheric $CO_2$ on plant ecophysiology (17). However, easily discernible effects on animals, plants, and insects arising from rapid global warming in the past three decades have exposed the overriding role of climate change.

A sudden widespread decline of frogs, with extinction of entire mountain-restricted species attributed to global warming (81, 82), provided a dramatic awakening. There are multiple causes of the detailed processes involved in global amphibian declines and extinctions (83, 84), but there is agreement that global warming is a key contributor and portends a planetary-scale mass extinction in the making unless humanity takes prompt action to stabilize climate while also fighting biodiversity's other threats (85).

Mountain-restricted and polar-restricted species are particularly vulnerable. As isotherms move up the mountainside and poleward, so does the climate zone in which a given species can survive. If global warming continues unabated, many of these species will be effectively pushed off the planet. There are already reductions in the population and health of Arctic species in the southern parts of the Arctic, Antarctic species in the northern parts of the Antarctic, and alpine species worldwide (17).

A critical factor for survival of some Arctic species is retention of all-year sea ice. Continued growth of fossil fuel emissions will cause loss of all Arctic summer sea ice within several decades. In contrast, the scenario in Fig.5a, with global warming peaking just over 1°C and then declining slowly, should allow summer sea ice to survive and then gradually increase to levels representative of recent decades.

The threat to species survival is not limited to mountain and polar species. Plant and animal distributions are a reflection of the regional climates to which they are adapted. Although species attempt to migrate in response to climate change, their paths may be blocked by human-constructed obstacles or natural barriers such as coast lines. As the shift of climate zones becomes comparable to the range of some species, less mobile species can be driven to extinction. Because of extensive species interdependencies, this can lead to mass extinctions.

IPCC (74) reviewed studies relevant to estimating eventual extinctions. They estimate that if global warming exceeds 1.6°C above preindustrial, 9-31 percent of species will be committed to extinction. With global warming of 2.9°C, an estimated 21-52 percent of species will be committed to extinction.

Mass extinctions occurred several times in Earth's history (86), often in conjunction with rapid climate change. New species evolved over millions of years, but those time scales are almost beyond human comprehension. If we drive many species to extinction we will leave a more desolate planet for our children, grandchildren, and more generations than we can imagine.

**Coral Reef Ecosystems.** Coral reefs are the most biologically diverse marine ecosystem, often described as the rainforests of the ocean. Over a million species, most not yet described (87), are estimated to populate coral reef ecosystems generating crucial ecosystem services for at least 500 million people in tropical coastal areas. These ecosystems are highly vulnerable to the combined effects of ocean acidification and warming.

Acidification arises as the ocean absorbs $CO_2$, producing carbonic acid (88). Geochemical records show that ocean pH is already outside its range of the past several million years (89, 90). Warming causes coral bleaching, as overheated coral expel symbiotic algae and become vulnerable to disease and mortality (91). Coral bleaching and slowing of coral calcification already are causing mass mortalities, increased coral disease, and reduced reef carbonate accretion, thus disrupting coral reef ecosystem health (14, 92).

Local human-made stresses add to the global warming and acidification effects, all of these driving a contraction of 1-2% per year in the abundance of reef-building corals (13). Loss of the three-dimensional coral reef frameworks has consequences for the millions of species that depend on them. Loss of these frameworks also has consequences for the important roles that coral reefs play in supporting fisheries and protecting coastlines from wave stress. Consequences of lost coral reefs can be economically devastating for many nations, especially in combination with other impacts such as sea level rise and intensification of storms.

**Climate Extremes.** Extremes of the hydrologic cycle are expected to intensify in a warmer world. A warmer atmosphere holds more moisture, so heavy rains become more intense, bringing more frequent and intense flooding. Higher temperatures, on the other hand, increase evaporation and intensify droughts, as does expansion of the subtropics with global warming. Heat waves lasting for weeks have a devastating impact on human health: the European heat wave of summer 2003 caused over 70,000 excess deaths (93). This heat record for Europe was surpassed already in 2010 (94). The number of extreme heat waves has increased several-fold due to global warming (18, 95) and will be multiplied further if temperatures continue to rise.

IPCC reports (2, 74) confirm that precipitation has generally increased over land poleward of the subtropics and decreased at lower latitudes. Unusually heavy precipitation events have increased in Europe, North America, Southeast Asia and Australia. Droughts are more common, especially in the tropics and subtropics.

Glaciers are in near-global retreat (74). Loss of glaciers can degrade the supply of fresh water to millions of people (7). Increased winter snowfall with a warmer moister atmosphere will tend to increase spring flooding but leave rivers drier during the driest months.

**Human Health.** Climate change causes a variety of human health impacts, with children especially vulnerable. These include food shortages, polluted air, and contaminated or scarce supplies of water, along with an expanding area of vectors causing infectious diseases and more intensely allergenic plants. World health experts have concluded with "very high confidence" that climate change already contributes to the global burden of disease and premature death (74).

IPCC (74) projects the following trends, if $CO_2$ emissions and global warming continue to increase, where only trends assigned very high confidence or high confidence are included: (i) increased malnutrition and consequent disorders, including those related to child growth and development, (ii) increased death, disease and injuries from heat waves, floods, storms, fires and droughts, (iii) increased cardio-respiratory morbidity and mortality associated with ground-level ozone. While IPCC also projects fewer deaths from cold, this positive effect is far outweighed by the negative ones.

With the growing awareness of the consequences of human-caused climate change, adults and children especially are susceptible to a range of anxiety and depressive disorders. Children cannot avoid hearing that the window of opportunity to act in time to avoid dramatic climate impacts is closing, and that their future and that of other species is at stake. While the psychological health of our children needs to be protected, denial of the truth exposes them to even greater risk.

The Supporting Information has further discussion of health impacts of climate change.

## Implications for Humanity

Fossil fuel emissions to date are a small fraction of potential emissions from known reserves and potentially recoverable resources (Fig. P1). Although there are uncertainties in reserves and resources, ongoing fossil fuel subsidies and continuing technological advances ensure that more and more of these fuels will be economically recoverable.

Burning all fossil fuels would create a very different planet than the one that humanity knows. The paleoclimate record and ongoing climate change make it clear that the climate system would be pushed beyond tipping points, setting in motion irreversible changes, including ice sheet disintegration with a continually adjusting shoreline, extermination of a substantial fraction of species on the planet, and increasingly devastating regional climate extremes.

Initiation of phase-out of fossil fuel emissions is urgent. $CO_2$ from fossil fuel use stays in the surface climate system for millennia. Thus continued high emissions would leave young people and future generations with an enormous clean-up job. The task of extracting $CO_2$ from the air is so great that success is uncertain at best, raising the likelihood of a spiral into climate catastrophes and efforts to "geo-engineer" restoration of planetary energy balance.

Most proposed schemes to artificially restore Earth's energy balance aim to reduce solar heating, e.g., by maintaining a haze of stratospheric particles that reflect sunlight to space. Such attempts to mask one pollutant with another pollutant almost inevitably would have unintended consequences. Moreover, schemes that do not remove $CO_2$ would not avert ocean acidification.

The implication is that the world must move expeditiously to carbon-free energies and energy efficiency, leaving most remaining fossil fuels in the ground. Yet transition to a post-fossil fuel world of clean energies will not occur as long as fossil fuels are the cheapest energy. Fossil fuels are cheap only because they are subsidized and do not pay their costs to society. Air and water pollution from fossil fuel extraction and use have high costs in human health, food production, and natural ecosystems, costs borne by the public. Huge costs of climate change and ocean acidification also are borne by the public, especially young people and future generations.

Thus the essential underlying policy, albeit not sufficient, is for emissions of $CO_2$ to come with a price that allows these costs to be internalized within the economics of energy use. The price should rise over decades to enable people and businesses to efficiently adjust their lifestyles and investments to minimize costs. The right price for carbon and the best mechanism for carbon pricing are more matters of practicality than of economic theory.

Economic analyses indicate that a carbon price fully incorporating environmental and climate damage would be high (96). The cost of climate change is uncertain to a factor of 10 or more and could be as high as ~$1000/tCO_2$ (97). While the imposition of such a high price on carbon emissions is outside the realm of short-term political feasibility, a price of that magnitude is not required to engender a large change in emissions trajectory.

An economic analysis indicates that a tax beginning at $15/tCO_2$ and rising $10/tCO_2$ each year would reduce emissions in the U.S. by 30% within 10 years (98). Such a reduction is more than 10 times as great as the carbon content of tar sands oil carried by the proposed Keystone XL pipeline (830,000 barrels/day) (99). Reduced oil demand would be nearly six times the pipeline capacity (98), thus rendering it superfluous.

Relative merits of a carbon tax and cap-and-trade have long been debated (100). A cap-and-trade system for $CO_2$ emissions was implemented in Europe, but not in the U.S., where opponents of any action on climate won the political battle by branding cap-and-trade as a devious new tax. However, a gradually rising fee on carbon emissions collected from fossil fuel companies with proceeds fully distributed to the public, was praised by the policy director of Republicans for Environmental Protection (101) as: "Transparent. Market-based. Does not enlarge government. Leaves energy decisions to individual choices… Sounds like a conservative climate plan."

A rising carbon emissions price is the *sine qua non* for fossil fuel phase out. However, it is not sufficient. Governments also should encourage investment in energy R&D and drive energy and carbon efficiency standards for buildings, vehicles and other manufactured products. Investment in global climate monitoring systems and support for climate mitigation and adaptation in undeveloped countries are also needed.

Despite evidence that a rising carbon price and these supplementary actions would drastically shrink demand for fossil fuels, governments and businesses are rushing headlong into expanded extraction and use of all fossil fuels. How is it possible that large human-driven climate change is unfolding virtually unimpeded, despite scientific understanding of likely consequences? Would not governments – presumably instituted for the protection of all citizens – have stepped in to safeguard the future of young people? A strong case can be made that the absence of effective leadership in most nations is related to the undue sway of special financial interests on government policies aided by pervasive public relations efforts by organizations that profit from the public's addiction to fossil fuels and wish to perpetuate that dependence (102, 103).

A situation in which scientific evidence cries out for action, but a political response is impeded by the financial power of special interests, suggests the possibility of an important role for the judiciary system. Indeed, in some nations the judicial branch of government may be able to require the executive branch to present realistic plans to protect the rights of the young (104). Such a legal case for young people should demand plans for emission reductions that are consistent with what the science shows is required to stabilize climate. Judicial recognition of both the exigency of the climate problem and the rights of young people, we believe, will help draw attention to the need for a rapid change of direction.

Nevertheless, fundamental change is unlikely without public support. Gaining that support requires widespread recognition that a prompt orderly transition to the post fossil fuel world, via a rising price on carbon emissions, is technically feasible and may even be economically beneficial apart from the benefits to climate.

The most basic matter is not one of economics, however. It is a matter of morality – a matter of intergenerational justice. As with the earlier great moral issue of slavery, an injustice done by one race of humans to another, so the injustice of one generation to others must stir the public's conscience to the point of action. Until there is a sustained and growing public involvement, it is unlikely that the needed fundamental change of direction can be achieved.

A broad public outcry may seem unlikely given the enormous resources of the fossil fuel industry, which allows indoctrination of the public with the industry's perspective. The merits of coal, of oil from tar sands and the deep ocean, of gas from hydrofracking are repeatedly extolled, all of these supposedly to be acquired with utmost care of the environment. Potential climate concerns are addressed, if at all, by discrediting climate science and scientists (102).

Yet human cultures have long revered the environment and other life on the planet, and an obligation to future generations is broadly recognized. Religious leaders have expressed support for ameliorating the causes of human-made climate change, in messages ranging from general statements by the Catholic Pope (105) to specific endorsement by the Buddhist Dalai Lama (106) of the target to reduce atmospheric $CO_2$ below 350 ppm. In our Supporting Information, the Executive Coordinator of the National Religious Coalition on Creation Care describes support for actions to stem climate change by an array of religions in the United States spanning Evangelical, mainline Protestant, Catholic, Jewish, and Eastern Orthodox faiths.

Indigenous people and people in developing countries have done little to cause climate change but will likely suffer some of the worst consequences. Many are resisting exploitation of their land and peacefully demanding policy changes. Indigenous people, farmers, scientists, environmentalists and other members of the public have peacefully demonstrated against new fossil fuel developments such as the tar sands Keystone pipeline (107) and hydrofracking in the Delaware River Basin.

Considering the stakes involved, it is disquieting that young people have not become more involved in the issue of the planet's future and more insistent on intergenerational justice. The tentative efforts to pursue legal redress, for which our present paper provides scientific rationale and quantification, are an effort of adults on behalf of young people. In the case of the very young, their inactivity is understandable. College-educated youth are equipped to understand the predicament and articulate their case, but their numbers so far have been too modest for their voice to compete against special financial interests (102, 103).

Yet it is possible to imagine a scenario in which a social tipping point is reached and the world begins to rapidly phase out fossil fuel emissions. When public concern reaches a high level, some influential leaders of the energy industry, well aware of the moral issue, could join the campaign to phase out fossil fuel emissions. Many business leaders recognize the merits of a rising price on carbon emissions, and likely would be supportive of such an approach once they realize that large rapid emission reductions are essential. Given the relative ease with which a carbon price can be made international (100), a rapid phasedown of emissions may be feasible. As fossil fuels are made to pay their costs to society, energy efficiency and clean energies could themselves reach a tipping point where they begin to be rapidly adopted.

Can the human tipping point be reached before the climate system passes a point of no return? What we have shown in this paper is that time is rapidly running out. The era of doubts, delays and denial, of ineffectual half-measures, must end. The period of consequences is beginning. If we fail to stand up now and demand a change of course, the blame will fall on us, the current generation of adults. Our parents did not know that their actions could harm future generations. We will only be able to pretend that we did not know. And that is unforgiveable.

**Acknowledgements.** This paper is dedicated to Paul Epstein, a fervent defender of the health of humans and the environment, who graciously provided important inputs to this paper in the spring and summer of 2011 while battling late stages of non-Hodgkin's lymphoma. We thank Inez Fung and Charles Komanoff for perceptive helpful reviews and Mark Chandler, Bishop Dansby, Ian Dunlop, Dian Gaffen Seidel, Edward Greisch, Fred Hendrick, Tim Mock, Ana Prados, and Rob Socolow for helpful suggestions on a draft of the paper.

References

1. Intergovernmental Panel on Climate Change (IPCC), 2007: *Climate Change 2007: The Physical Science Basis.* Solomon, S*., et al.* eds., Cambridge University Press, 996 pp

2. Intergovernmental Panel on Climate Change (IPCC), 2001: *Climate Change 2001: The Scientific Basis.* Houghton, J. T*., et al.* eds., Cambridge University Press, 996 pp

3. Schneider, S.H. and Mastrandrea, M.D., 2005: Probabilistic assessment "dangerous" climate change and emissions pathways. *Proc Nat Acad Sci*, **102**, 15728-15735.

4. Rampal, P., Weiss, J., Dubois, C., and Campin, J.M., 2011: IPCC climate models do not capture Arctic sea ice drift acceleration: Consequences in terms of projected sea ice thinning and decline. *J Geophys Res*, **116**, C00D07.

5. Velicogna, I., 2009: Increasing rates of ice mass loss from the Greenland and Antarctic ice sheets revealed by GRACE. *Geophys Res Lett*, **36**, L19503.

6. Rignot, E., Velicogna, I., van den Broeke, M.R., Monaghan, A., and Lenaerts, J., 2011: Acceleration of the contribution of the Greenland and Antarctic ice sheets to sea level rise. *Geophys Res Lett*, **38**, L05503-L05508.

7. Barnett, T.P., *et al.*, 2008: Human-induced changes in the hydrology of the western United States. *Science*, **319**, 1080-1083.

8. Kaser, G., Grosshauser, M., and Marzeion, B., 2010: Contribution potential of glaciers to water availability in different climate regimes. *Proc Nat Acad Sci*, **107**, 20223-20227.

9. Held, I.M. and Soden, B.J., 2006: Robust responses of the hydrological cycle to global warming. *J Clim*, **19**, 5686-5699.

10. Seidel, D.J., Fu, Q., Randel, W.J., and Reichler, T.J., 2008: Widening of the tropical belt in a changing climate. *Nat Geosci*, **1**, 21-24.

11. Davis, S.M. and Rosenlof, K.H., 2011: A multi-diagnostic intercomparison of tropical width time series using reanalyses and satellite observations. *J. Clim.*, in press.

12. Westerling, A.L., Hidalgo, H.G., Cayan, D.R., and Swetnam, T.W., 2006: Warming and earlier spring increase western US forest wildfire activity. *Science*, **313**, 940-943.

13. Bruno, J.F. and Selig, E.R., 2007: Regional decline of coral cover in the Indo-Pacific: timing, extent, and subregional comparisons. *Plos One*, **2**, e711.

14. Hoegh-Guldberg, O., *et al.*, 2007: Coral reefs under rapid climate change and ocean acidification. *Science*, **318**, 1737-1742.

15. Veron, J.E., *et al.*, 2009: The coral reef crisis: The critical importance of < 350 ppm CO(2). *Mar Pollut Bull*, **58**, 1428-1436.

16. Parmesan, C. and Yohe, G., 2003: A globally coherent fingerprint of climate change impacts across natural systems. *Nature*, **421**, 37-42.

17. Parmesan, C., 2006: Ecological and evolutionary responses to recent climate change. *Annu. Rev. Ecol. Syst.*, **37**, 637-669.

18. Rahmstorf, S. and Coumou, D., 2011: Increase of extreme events in a warming world. *Proc Nat Acad Sci*, **108**, 17905-17909.

19. Zachos, J., Pagani, M., Sloan, L., Thomas, E., and Billups, K., 2001: Trends, rhythms, and aberrations in global climate 65 Ma to present. *Science*, **292**, 686-693.

20. Hansen, J., *et al.*, 2008: Target Atmospheric $CO_2$: Where Should Humanity Aim? *The Open Atmospheric Science Journal*, **2**, 217-231.

21. Hansen, J.E. and Sato, M., Berger, A., Mesinger, F., and Sijacki, D., 2012: *Paleoclimate implications for human-made climate change.* . Springer, ~350 pp.

22. Hansen, J., *et al.*, 2007: Climate change and trace gases. *Philos Trans R Soc A*, **365**, 1925-1954.

23. Kohler, P., *et al.*, 2010: What caused Earth's temperature variations during the last 800,000 years? Data-based evidence on radiative forcing and constraints on climate sensitivity. *Quat. Sci. Rev.*, **29**, 129-145.

24. Rohling, E.J., M. Medina-Elizalde, J.G. Shepherd, M. Siddall, and J.D. Stanford, 2011: Sea surface and high-latitude temperature sensitivity to radiative forcing of climate over several glacial cycles. *J Clim*, in press.

25. Pagani, M., Liu, Z.H., LaRiviere, J., and Ravelo, A.C., 2010: High Earth-system climate sensitivity determined from Pliocene carbon dioxide concentrations. *Nat Geosci*, **3**, 27-30.

26. Rohling, E.J., *et al.*, 2008: High rates of sea-level rise during the last interglacial period. *Nat Geosci*, **1**, 38-42.

27. Thompson, W.G., Curran, H.A., Wilson, M.A., and White, B., 2011: Sea-level oscillations during the last interglacial highstand recorded by Bahamas corals. *Nat Geosci*, **4**, 684-687.

28. Hearty, P.J., Hollin, J.T., Neumann, A.C., O'Leary, M.J., and McCulloch, M., 2007: Global sea-level fluctuations during the Last Interglaciation (MIS 5e). *Quat. Sci. Rev.*, **26**, 2090-2112.

29. Raymo, M.E., Mitrovica, J.X., O'Leary, M.J., DeConto, R.M., and Hearty, P.L., 2011: Departures from eustasy in Pliocene sea-level records. *Nat Geosci*, **4**, 328-332.

30. Naish, T.R. and Wilson, G., 2009: Constraints on the amplitude of Mid-Pliocene (3.6-2.4 Ma) eustatic sea-level fluctuations from the New Zealand shallow-marine sediment record. *Phil Trans R Soc A* **367**, 169-187.

31. Hill, D.J., Haywood, D.M., Hindmarsh, R.C.M., and Valdes, P.J. (2007) Characterizing ice sheets during the Pliocene: evidence from data and models. *Deep-Time Perspectives on Climate Change: Marrying the Signal from Computer Models and Biological Proxies*, Williams, M., Haywood, A. M., Gregory, J., and Schmidt, D. N. eds, Micropalaeont Soc Geol Soc London, pp 517-538.

32. Dwyer, G.S. and Chandler, M.A., 2009: Mid-Pliocene sea level and continental ice volume based on coupled benthic Mg/Ca palaeotemperatures and oxygen isotopes. *Phil Trans Royal Soc A*, **367**, 157-168.

33. Hansen, J., Mki. Sato, P. Kharecha, and Schuckmann, K.v., 2011: Earth's Energy Imbalance and Implications. *Atmos Chem Phys*, **11**, 1-29.

34. Roemmich, D. and Gilson, J., 2009: The 2004-2008 mean and annual cycle of temperature, salinity, and steric height in the global ocean from the Argo Program. *Prog Oceanogr*, **82**, 81-100.

35. Lyman, J.M., *et al.*, 2010: Robust warming of the global upper ocean. *Nature*, **465**, 334-337.

36. Barker, P.M., Dunn, J.R., Domingues, C.M., and Wijffels, S.E., 2011: Pressure Sensor Drifts in Argo and Their Impacts. *J Atmos Ocean Tech*, **28**, 1036-1049.

37. von Schuckmann, K. and LeTraon, P.-Y., 2011: How well can we derive Global Ocean Indicators from Argo data? *Ocean Sci*, **7**, 783-391.

38. Frohlich, C. and Lean, J., 1998: The Sun's total irradiance: Cycles, trends and related climate change uncertainties since 1976. *Geophys Res Lett*, **25**, 4377-4380.

39. Hansen, J., Sato, M., Ruedy, R., Lacis, A., and Oinas, V., 2000: Global warming in the twenty-first century: An alternative scenario. *Proc Nat Acad Sci*, **97**, 9875-9880.

40. Archer, D., 2007: Methane hydrate stability and anthropogenic climate change. *Biogeosciences*, **4**, 521-544.

41. Joos, F., *et al.*, 1996: An efficient and accurate representation of complex oceanic and biospheric models of anthropogenic carbon uptake. *Tellus (B Chem. Phys. Meteorol.)*, **48**, 397-417.

42. Kharecha, P.A. and Hansen, J.E., 2008: Implications of "peak oil" for atmospheric $CO_2$ and climate. *Global Biogeochem Cy*, **22**, GB3012.

43. Stocker, B.D., Strassmann, K., and Joos, F., 2011: Sensitivity of Holocene atmospheric $CO_2$ and the modern carbon budget to early human land use: analyses with a process-based model. *Biogeosciences*, **8**, 69-88.

44. Hillel, D. and Rosenzweig, C., eds. 2011: *Handbook of Climate Change and Agroecosystems: Impacts, Adaptation and Mitigation* (Imperial College Press, London).

45. Lamb, D., 2011: *Regreening the Bare Hills.* Springer, New York, 547 pp.

46. Smith, P., *et al.*, 2008: Greenhouse gas mitigation in agriculture. *Philos T R Soc B*, **363**, 789-813.

47. Smith, P., 2012: Agricultural greenhouse gas mitigation potential globally, in Europe and in the UK: what have we learned in the last 20 years? . *Global Change Biol.*, **18**, 35-43.

48. Smith, P., *et al.*, 2010: Competition for land. *Philos T R Soc B*, **365**, 2941-2957.

49. Bala, G., *et al.*, 2007: Combined climate and carbon-cycle effects of large-scale deforestation. *Proc Natl Acad Sci USA*, **104**, 6550-6555.

50. Bonan, G.B., 2008: Forests and climate change: Forcings, feedbacks, and the climate benefits of forests. *Science*, **320**, 1444-1449.

51. Swann, A.L.S., Fung, I.Y., and Chiang, J.C.H., 2011: Mid-latitude afforestation shifts general circulation and tropical precipitation. *Proc Natl Acad Sci (Early Edition)*, www.pnas.org/cgi/doi/10.1073/pnas.1116706108.

52. Pan, Y.D., *et al.*, 2011: A Large and Persistent Carbon Sink in the World's Forests. *Science*, **333**, 988-993.

53. Zomer, R.J., Trabucco, A., Bossio, D.A., and Verchot, L.V., 2008: Climate change mitigation: A spatial analysis of global land suitability for clean development mechanism afforestation and reforestation. *Agr Ecosyst Environ*, **126**, 67-80.

54. Tilman, D., Hill, J., and Lehman, C., 2006: Carbon-negative biofuels from low-input high-diversity grassland biomass. *Science*, **314**, 1598-1600.

55. Fargione, J., Hill, J., Tilman, D., Polasky, S., and Hawthorne, P., 2008: Land clearing and the biofuel carbon debt. *Science*, **319**, 1235-1238.

56. Searchinger, T.*, et al.*, 2008: Use of US croplands for biofuels increases greenhouse gases through emissions from land-use change. *Science*, **319**, 1238-1240.

57. Stehfest, E.*, et al.*, 2009: Climate benefits of changing diet. *Clim Chg*, **95**, 83-102.

58. United Nations Environment Programme (UNEP), 2010: *Assessing the Enviromental Impacts of Consumption and Production: Priority Products and Materials.* Hertwich, E.*, et al.*, www.unep.org/resourcepanel/Publications/PriorityProducts/tabid/56053/Default.aspx.

59. The Royal Society, 2009: *Geoengineering the climate: science, governance and uncertainty.*

60. Keith, D.W., Ha-Duong, M., and Stolaroff, J.K., 2006: Climate strategy with $CO_2$ capture from the air. *Clim Chg*, **74**, 17-45.

61. American Physical Society, 2011: Direct Air Capture of $CO_2$ with Chemicals: A Technology Assessment for the APS Panel on Public Affairs. : http://www.aps.org/policy/reports/assessments/upload/dac2011.pdf accessed Jan 11, 2012.

62. Hansen, J. and Sato, M., 2004: Greenhouse gas growth rates. *Proc Nat Acad Sci*, **101**, 16109-16114.

63. Lenton, T.M.*, et al.*, 2008: Tipping elements in the Earth's climate system. *Proc. Natl. Acad. Sci*, **105**, 1786-1793.

64. Max, M.D., 2003: *Natural Gas Hydrate in Oceanic and Permafrost Environments.* Kluwer Academic Publishers,, ISBN 0-7923-6606-9.

65. Kvenvolden, K.A., 1993: Gas Hydrates - Geological Perspective and Global Change. *Rev Geophys*, **31**, 173-187.

66. Walter, K., Zimov, S., Chanton, J., Verbyla, D., and Chapin, F., 2006: Methane bubbling from Siberian thaw lakes as a positive feedback to climate warming. *Nature*, **443**, 71-75.

67. Shakhova, N.*, et al.*, 2010: Extensive Methane Venting to the Atmosphere from Sediments of the East Siberian Arctic Shelf. *Science*, **327**, 1246-1250.

68. Lourens, L.J.*, et al.*, 2005: Astronomical pacing of late Palaeocene to early Eocene global warming events. *Nature*, **435**, 1083-1087.

69. Zachos, J.C., Dickens, G.R., and Zeebe, R.E., 2008: An early Cenozoic perspective on greenhouse warming and carbon-cycle dynamics. *Nature*, **451**, 279-283.

70. Lunt, D.J., *et al.*, 2011: A model for orbital pacing of methane hydrate destabilization during the Palaeogene. *Nat Geosci*, **4**, 775-778.

71. Muhs, D.R., Simmons, K.R., Schumann, R.R., and Halley, R.B., 2011: Sea-level history of the past two interglacial periods: new evidence from U-series dating of reef corals from south Florida. *Quat. Sci. Rev.*, **30**, 570-590.

72. Hearty, P.J. and Neumann, A.C., 2001: Rapid sea level and climate change at the close of the Last Interglaciation (MIS 5e): evidence from the Bahama Islands. *Quat. Sci. Rev.*, **20**, 1881-1895.

73. Rignot, E., *et al.*, 2008: Recent Antarctic ice mass loss from radar interferometry and regional climate modelling. *Nat Geosci*, **1**, 106-110.

74. Hansen, J.E., 2007: Scientific reticence and sea level rise. *Env Res Lett*, **2**, 024002.

75. Hansen, J.E., 2005: A slippery slope: How much global warming constitutes "dangerous anthropogenic interference"? *Clim Chg*, **68**, 269-279.

76. Levi, B.G., 2008: Trends in the hydrology of the western US bear the imprint of manmade climate change. *Physics Today*, **61**, 16-18.

77. Hansen, J., *et al.*, 2006: Global temperature change. *Proc Nat Acad Sci*, **103**, 14288-14293.

78. Burrows, M.T., *et al.*, 2011: The Pace of Shifting Climate in Marine and Terrestrial Ecosystems. *Science*, **334**, 652-655.

79. Seimon, T.A., *et al.*, 2007: Upward range extension of Andean anurans and chytridiomycosis to extreme elevations in response to tropical deglaciation. *Global Change Biol.*, **13**, 288-299.

80. Hoegh-Guldberg, O. and Bruno, J.F., 2010: The Impact of Climate Change on the World's Marine Ecosystems. *Science*, **328**, 1523-1528.

81. Pounds, J.A., Fogden, M.P.L., and Campbell, J.H., 1999: Biological response to climate change on a tropical mountain. *Nature*, **398**, 611-615.

82. Pounds, J.A., *et al.*, 2006: Widespread amphibian extinctions from epidemic disease driven by global warming. *Nature*, **439**, 161-167.

83. Alford, R.A., Bradfield, K.S., and Richards, S.J., 2007: Ecology: Global warming and amphibian losses. *Nature*, **447**, E3-E4.

84. Rosa, I.D., Simoncelli, F., Fagotti, A., and Pascolini, R., 2007: Ecology: The proximate cause of frog declines? *Nature*, **447**, E4-E5.

85. Pounds, J.A., *et al.*, 2007: Ecology - Pounds et al. reply. *Nature*, **447**, E5-E6.

86. Raup, D.M. and Sepkoski, J.J., 1982: Mass Extinctions in the Marine Fossil Record. *Science*, **215**, 1501-1503.

87. Reaka-Kudla, M.L. (1997) Global biodiversity of coral reefs: a comparison with rainforests. *Biodiversity II: Understanding and Protecting Our Biological Resources*, Reaka-Kudla, M. L., And Wilson, D.E. ed, Joseph Henry Press, Vol II, p 551.

88. Caldeira, K. and Wickett, M.E., 2003: Oceanography: Anthropogenic carbon and ocean pH. *Nature*, **425**, 365-365.

89. 2005: *Ocean acidification due to increasing atmospheric carbon dioxide.* London Royal Society, Raven, J*., et al.*,

90. Pelejero, C., Calvo, E., and Hoegh-Guldberg, O., 2010: Paleo-perspectives on ocean acidification. *Trends Ecol Evol*, **25**, 332-344.

91. Hoegh-Guldberg, O., 1999: Climate change, coral bleaching and the future of the world's coral reefs. *Mar Freshwater Res*, **50**, 839-866.

92. De'ath, G., Lough, J.M., and Fabricius, K.E., 2009: Declining Coral Calcification on the Great Barrier Reef. *Science*, **323**, 116-119.

93. Robine, J.M.*, et al.*, 2008: Death toll exceeded 70,000 in Europe during the summer of 2003. *C. R. Biol.*, **331**, 171-175.

94. Barriopedro, D., Fischer, E.M., Luterbacher, J., Trigo, R., and Garcia-Herrera, R., 2011: The Hot Summer of 2010: Redrawing the Temperature Record Map of Europe. *Science*, **332**, 220-224.

95. Stott, P.A., Stone, D.A., and Allen, M.R., 2004: Human contribution to the European heatwave of 2003. *Nature*, **432**, 610-614.

96. Stern, N., 2007: *Stern Review on the Economics of Climate Change* Cambridge Univ. Press, Cambridge, UK.

97. Ackerman, F., DeCanio, S., Howarth, R., and Sheeran, K., 2009: Limitations of integrated assessment models of climate change. *Clim Chg*, **95**, 297-315.

98. Komanoff, C., 2011: 5-Sector Carbon Tax Model: http://www.komanoff.net/fossil/CTC_Carbon_Tax_Model.xls accessed December 25, 2011.

99. United States Department of State, 2011: Final Environmental Impact Statement: http://www.keystonepipeline-xl.state.gov/clientsite/keystonexl.nsf/Fact%20Sheet.pdf?OpenFileResource accessed 28 December 2011.

100. Hsu, S.-L., 2011: *The Case for a Carbon Tax.* Island Press, Washington.

101. Dipeso, J., 2010: Jim Hansen's conservative climate plan, blog post at Republican's for Environmental Protection, October 11, 2010: http://www.rep.org/opinions/weblog/weblog10-10-11.html accessed August 26, 2011.

102. Oreskes, N. and Conway, E.M., 2010: *Merchants of Doubt: How a Handful of Scientists Obscured the Truth on Issues from Tobacco Smoke to Global Warming.* Bloomsbury Press, 355 pp., merchantsofdoubt.org.

103. Hansen, J., 2009: *Storms of My Grandchildren.* Bloomsbury, New York, 304 pp.

104. Wood, M.C. (2009): Atmospheric Trust Litigation, *Adjudicating Climate Change: Sub-National, National, And Supra-National Approaches.* Burns, W. C. G. Osofsky, H. M. eds., Cambridge Univ. Press, , pp. 99-125, available at http://www.law.uoregon.edu/faculty/mwood/docs/atmospheric.pdf.

105. Thavis, J., 2011: Pope urges international agreement on climate change, November 28, 2011: http://www.catholicnews.com/data/stories/cns/1104646.htm accessed December 31, 2011.

106. Dalai Lama (2009) Endorsement of a safe level for atmospheric carbon dioxide. *A Buddhist Response to the Climate Emergency*, Stanley, J., Loy, D. R., and Dorje, G. eds, Wisdom Publications, pp 270-273.

107. McKibben, B., The Keystone Pipeline Revolt: Why Mass Arrests are Just the Beginning, *Rolling Stone.* http://www.rollingstone.com/politics/news/the-keystone-pipeline-revolt-why-mass-arrests-are-just-the-beginning-20110928 accessed Dec 25, 2011.

108. Hansen, J., Ruedy, R., Sato, M., and Lo, K., 2010: Global Surface Temperature Change. *Rev Geophys*, **48**, RG4004.

# Supplementary Material

**Sea Level Change**

Recent estimates of sea level rise by 2100 are around 1 m (1, 2). Ice-dynamics studies estimate that rates of sea-level rise of 0.8 to 2 m per century are feasible (3) and Antarctica alone could contribute up to 1.5 m per century (4). Hansen (5, 6) has argued that BAU $CO_2$ emissions produce a climate forcing so much larger than any experienced in prior interglacial periods that a non-linear ice sheet response with multi-meter sea level rise could occur this century.

Accurate measurements of ice sheet mass loss may provide the best means to detect nonlinear change. The GRACE satellite, measuring Earth's gravitational field since 2003, reveals that the Greenland ice sheet is losing mass at an accelerating rate, now more than 200 $km^3$/year, and Antarctica is losing more than 100 $km^3$/year (7, 8). However, the present rate of sea level rise, 3 cm/decade, is moderate, and the ice sheet mass balance record is too short to determine whether we have entered a period of continually accelerating ice loss.

Satellite observations show that the Greenland surface area with summer melting has increased over the period of record, which extends from the late 1970s (9, 10). A destabilizing mechanism of possibly greater concern is melting of ice shelves, the tongues of ice that extend into the oceans, buttress the ice sheets, and limit the rate of discharge of ice into the ocean. Ocean warming is causing shrinkage of ice shelves around Greenland and Antarctica (11).

Most of the West Antarctic ice sheet, which alone could raise sea level by 3-5 meters, rests on bedrock below sea level, making that ice sheet vulnerable to rapid change. Parts of the larger East Antarctic ice sheet are also vulnerable. Satellite gravity and radar altimetry reveal that the Totten Glacier of East Antarctica, fronting a large ice mass grounded below sea level, is beginning to lose mass (12).

**Paleocene-Eocene Thermal Maximum (PETM)**

Rapid global warming of at least 5°C at the Paleocene-Eocene boundary (about 56 million years ago) provides valuable insights into the carbon cycle, climate system, and biotic responses to environmental change (13-15). The PETM event may be the closest analogy in Earth's history to the potential burning of most fossil fuels, because the PETM warming occurred in conjunction with injection of >3000 GtC of carbon into the climate system within ~5-10 thousand years (16, 17).

The most common interpretation is that the carbon originated mainly from melting of methane hydrates, because of the difficulty of identifying other large sources. One suggested alternative carbon source is release from Antarctic permafrost and peat (18). One reason for questioning the methane hydrate source, whether that source could be large enough given the warmer ocean at that time, has been affirmatively addressed (19). Regardless of the carbon source, PETM occurred during a 10-million year period of slow global warming driven by low-level volcanic carbon emissions, suggesting that the methane release may have initiated at a physical threshold, acting as a powerful feedback magnifying that warming. Support for the interpretation that the carbon release was an amplifying feedback is provided by evidence that several other PETM-like events in Earth's history (spikes in global warming and light-carbon sediments) were astronomically paced, i.e., they occurred during the warm phase of climate oscillations associated with perturbations of Earth's orbit (14).

The PETM witnessed global scale disruption of marine and terrestrial ecosystems with mass migration, temporary redistribution of many species toward higher latitude, and rapid evolution, particularly toward dwarfism of mammals, but with only minor extinctions (15). The evolution toward smaller body size may have been a result of a decline in biological productivity and food availability (20).

An important point is that the magnitude of the PETM carbon injection and warming is comparable to what will occur if humanity burns most of the fossil fuels, but the human-made warming is occurring 10-100 times faster. We have no empirical evidence on the ability of life on Earth to maintain itself during such a large, rapid climate change, with climate zones shifting much faster than species have ever experienced. The faster carbon addition also means that acidification and carbonate dissolution in the surface ocean would be more severe than that experienced by surface-dwelling organisms in the PETM.

**Human Health**

If fossil fuel emissions continue to increase rapidly, as in the business-as-usual scenarios of IPCC (21), substantial impacts of climate change on human health are likely. Some effects are already beginning to occur.

**Infectious Disease.** Increased temperature and flooding facilitate spread of infectious diseases by increasing the range and frequency of conditions favoring blood-sucking arthropods, such as mosquitoes, fleas, lice, biting flies, bugs and ticks. Warmer winters and polar amplification of warming are especially effective in increasing the range of these disease-bearing vectors. Tick-borne Lyme disease has expanded rapidly and become the most important vector-borne disease in the United States (http://www.columbia-lyme.org/research/abstracts.html).

**Crop Pests and Disease.** Warming fortifies pests and weakens hosts in forests, agricultural systems, and marine life. Warmer winters allow pine bark beetles to overwinter and expand their range, to the detriment of boreal forests. Climate trends also favor expansion of the Asian long-horned beetle and wooly adelgid to the detriment of trees in the Northeast United States. Warming increases the range of pests such as white flies, aphids and locust that damage crops, and it stimulates growth of agricultural weeds, leading to increased use of pesticides and herbicides that themselves are harmful to human health. Warming harms coral and other species hosted by coral reefs, and, along with excess nutrients from fertilizers, contributes to harmful algal blooms that cause dead zones in coastal waterways and estuaries (22).

**Heat Waves and Droughts.** Global warming, although "only" 0.8°C in the past century, is already sufficient to substantially increase the likelihood of extreme heat waves and droughts. The probability of occurrence of extreme anomalies as great as the Moscow heat wave in 2010 and the Texas/Oklahoma heat wave and drought of 2011 has increased by several times because of global warming (22), and the probability will increase even further if global warming continues to increase. Heat waves cause illness and death and also can lead to an increase in aggression, including violent assaults (23) and suicide (24).

**Food Insecurity.** Food supplies are compromised by increasing climate extremes, crop pests, and displacement of food crops by biofuel crops (21). Unusually extensive droughts and floods in 2010 caused widespread grain shortages and raised food prices, causing food riots in Uganda and Burkina Faso and likely contributing to political instability and uprisings in North Africa and the Middle East. Food shortages and price hikes contribute to malnutrition and poor health that increase vulnerability to infectious diseases, and also are frequently factors in conflicts and wars (21).

**Religions and Climate**

There is widespread support among religions for preserving climate and the environment. An indicative sample of religious statements follows.

**World Council of Churches.** At their meeting in Geneva Switzerland on 13-20 February 2008 the World Council of Churches called urgently for the churches to strengthen their moral stand in relationship to global warming and climate change, recalling its adverse effects on poor and vulnerable communities in various parts of the world, and encouraging the churches to reinforce their advocacy towards governments, NGOs, the scientific community and the business sector to intensify cooperation in response to global warming and climate change (http://nrccc.org/?page_id=40).

**Evangelicals.** Evangelical organizations are diverse, but leaders of American evangelical faiths have issued an evangelical call to action concerning climate change, recognizing a responsibility to offer biblically-based moral witness that helps shape public policy and contributes to the will-being of the world (http://nrccc.org/?page_id=42).

**Jewish Faith.** The Central Conference of American Rabbis adopted a resolution on climate change at their 116$^{th}$ annual convention in Houston Texas in March 2005, concluding that Jewish and secular moral principles imply an obligation to minimize climate change, to live within the ecological limits of Earth, and to not compromise the ecological or economic security of future generations (http://nrccc.org/?page_id=50).

**Orthodox Faith.** Patriarch Bartholomew II and the Standing Conference of Orthodox Bishops in America on 25 May 2007, in a "Global Climate Change: A Moral and Spiritual Challenge," concluded that care of the environment is an urgent issue, and that for humans to degrade the integrity of the Earth by causing changes in its climate is a sin (http://nrccc.org/?page_id=34).

**Catholic Faith.** Pope Benedict urged delegates at the United Nations climate conference to reach agreement on a responsible credible response to the the complex and disturbing effects of climate change (http://thinkprogress.org/romm/2011/11/29/377462).

**Southern Africa Religions.** Religious leaders from across South Africa met in Lusaka Zambia on 5-6 May 2011 to discuss climate change, recognizing the need for religions to help people retain a moral compass with a compassion for other living beings and the principle of justice (http://safcei.org/wp-content/uploads/2011/07/SA-Faith-Leaders-Declaration-09-05-2011.pdf).

**Canadian Interfaith.** Representatives of Canadian faith communities in 2011 stated their united conviction that the growing crisis of climate change needs to be met by solutions that draw upon the moral and spiritual resources of the world's religious traditions (http://www.cpj.ca/files/docs/Catalyst-Winter-2011.pdf).

## References

1. Vermeer, M. and Rahmstorf, S., 2009: Global sea level linked to global temperature. *Proc Nat Acad Sci,* **106**, 21527-21532.
2. Grinsted, A., Moore, J., and Jevrejeva, S., 2010: Reconstructing sea level from paleo and projected temperatures 200 to 2100 AD. *Clim Dyn,* **34**, 461-472.
3. Pfeffer, W. T., Harper, J. T., and O'Neel, S., 2008: Kinematic constraints on glacier contributions to 21st-century sea-level rise. *Science,* **321**, 1340-1343.

4. Turner J. et al. (eds.), 2009: *Antarctic Climate change and the environment: a contribution to the International Polar year 2007-2008*, Scientific Committee on Antarctic Research, Scott Polar Research Institute, Lensfield Road, Cambridge UK.
5. Hansen, J., 2005: A slippery slope: How much global warming constitutes "dangerous anthropogenic interference"? *Clim Chg,* **68**, 269-279.
6. Hansen, J., 2007: Scientific reticence and sea level rise. *Env Res Lett,* **2** 024002.
7. Sorensen, L. S. and Forsberg, R., 2010: Greenland Ice Sheet Mass Loss from GRACE Monthly Models. *Gravity, Geoid and Earth Observation,* **135**, 527-532, .
8. Rignot, E., Velicogna, I., van den Broeke, M. R., Monaghan, A., and Lenaerts, J., 2011: Acceleration of the contribution of the Greenland and Antarctic ice sheets to sea level rise. *Geophys Res Lett,* **38** L05503.
9. Steffen, K., Nghiem, S. V., Huff, R., and Neumann, G., 2004: The melt anomaly of 2002 on the Greenland Ice Sheet from active and passive microwave satellite observations. *Geophys Res Lett,* **31** L20402.
10. Tedesco, M.*, et al.*, 2011: The role of albedo and accumulation in the 2010 melting record in Greenland. *Env Res Lett,* **6** 014005.
11. Rignot, E. and Jacobs, S. S., 2002: Rapid bottom melting widespread near Antarctic ice sheet grounding lines. *Science,* **296**, 2020-2023.
12. Rignot, E.*, et al.*, 2008: Recent Antarctic ice mass loss from radar interferometry and regional climate modelling. *Nat Geosci,* **1**, 106-110.
13. Kennett, J. P., Stott, L. D.,, 1991: Abrupt deep sea warming, paleoceanographic changes, and benthic extinctions at the end of the Palaeocene:. *Nature,* **353**, 319-322.
14. Zachos, J., Pagani, M., Sloan, L., Thomas, E., and Billups, K., 2001: Trends, rhythms, and aberrations in global climate 65 Ma to present. *Science,* **292**, 686-693.
15. McInerney, F. A., Wing, S.L. , 2011: The Paleocene-Eocene Thermal Maximum-a perturbation of carbon cycle, climate, and biosphere with implications for the future. *Ann Rev Earth Plan Sci,* **39**, 489-516.
16. Zeebe, R. E., Zachos, J. C., and Dickens, G. R., 2009: Carbon dioxide forcing alone insufficient to explain Palaeocene-Eocene Thermal Maximum warming. *Nat Geosci,* **2**, 576-580.
17. Cui, Y.*, et al.*, 2011: Slow release of fossil carbon during the Palaeocene–Eocene Thermal Maximum. *Nat Geosci,* **4**, 481-485.
18. DeConto, R., Galeotti, S., Pagani, M., Tracy, D.M., Pollard, D., Beerling, D.J.. 2010: Hyperthermals and orbitally paced permafrost soil organic carbon dynamics. *Geophys Res Abstracts,* **13**, EGU2011-13580.
19. Gu, G., Dickens, G.R., Bhatnagar, G. Colwell, F.S., Hirasaki G.J., Chapman, W.G., 2011: Abundant Early Palaeogene marine gas hydrates despite warm deep-ocean temperatures. *Nat Geosci,* **4**, 848-851.
20. Chester, S., Bloch, J., Secord, R., Boyer, D., 2010: A new small-bodied species of Palaeonictis (Creodonta, Oxyaenidae) from the Paleocene-Eocene Thermal Maximum. *J Mamm Evol,* **17**, 227-243.
21. Intergovernmental Panel on Climate Change (IPCC), 2007: *Climate Change 2007, Impacts, Adaptation and Vulnerability*, M.L. Parry, E. A. ed., Cambridge Univ Press, 996 pp.
22. Diaz, R. J., Rosenberg, R., 2008: Spreading dead zones and consequences for marine ecosystems. *Science,* **321**, 926-928.
23. Bushman, B. J., Wang, M. C., and Anderson, C. A., 2005: Is the Curve Relating Temperature to Aggression Linear or Curvilinear? Assaults and Temperature in Minneapolis Reexamined. *J Personality, Social Psychology,* **89**, 62-66.
24. Page, L. A., Hajat, S. Kovats, R.S., 2007: Relationship between daily suicide counts and temperature in England and Wales *British J Psych,* **191**, 106-112.